\documentstyle[12pt,epsfig,hpatex]{article}

\begin{document}
\hfill ETH-TH/96-24\\[-12 pt]
\title{Cosmological Perturbations of Ultrarelativistic Plasmas}
\authors{Dominik J. Schwarz}
\address{Institut f\"ur
Theoretische Physik, ETH-H\"onggerberg, CH-8093 Z\"urich}
\abstract{
Scalar cosmological perturbations of a weakly self-interacting plasma
mixed with a perfect radiation fluid are investigated. Effects of this 
plasma are considered through order $\lambda^{3/2}$ of perturbative 
thermal-field-theory in the radiation dominated universe. 
The breakdown of thermal perturbation theory at vastly subhorizon scales is 
circumvented by a Pad\'e approximant solution. Compared to collisionless 
plasmas the phase speed and subhorizon damping of 
the plasma density perturbations are changed.
An example for a self-interacting thermal field is provided by the neutrinos
with effective 4-fermion interactions.
}

The evolution of cosmological perturbations depends on the matter content 
of the universe. In the radiation dominated epoch {\it photons, electrons}, 
and {\it baryons} are coupled strongly, hence an effective description 
as a perfect radiation fluid is possible. The energy density of the perfect
fluid is related to its pressure by the equation of state $\rho
= 3 p$. The cosmological perturbations \cite{cp} are determined by the
linearized Einstein equations together with the perturbed equation of state
$\delta\!\rho = c_s^2 \delta\!p$, where $c_s$ is the sound 
speed of the perfect radiation fluid; the anisotropic pressure vanishes.
{\it Neutrinos} interact weakly, therefore they decouple from the
strongly interacting matter at a temperature $\sim 1$ MeV 
and propagate freely thereafter. This 
almost collisionless gas of neutrinos may be described by the linearized
Einstein-Vlasov equations \cite{EV} or by a thermal-field-theoretic approach
\cite{Kraemmer}. In the latter the perturbed energy-momentum tensor is
calculated from the response of the ultrarelativistic plasma on a 
metric perturbation
\begin{equation}
\label{deltaT}
\delta(\sqrt{-g} T^{\mu\nu})(x) = \int_y {\delta^2 \Gamma \over
\delta\!g_{\mu\nu}(x) \delta\!g_{\alpha\beta}(y)} 
\delta\!g_{\alpha\beta}(y) \ ,
\end{equation}
where $\Gamma$ is the effective action of (thermal) matter. The Hubble rate 
$H \sim T^2/m_{\rm Planck}$ is much smaller than the temperature
$T$, thus momenta $k^{\rm phys}$ of cosmological interest are much smaller
than $T$ as well. The second variation of the effective action,
the gravitational polarization tensor,
is evaluated in the high temperature limit, $T\gg k^{\rm phys}$. 
The thermal-field-theoretic formulation goes beyond classical kinetic theory. 
Eq.~(\ref{deltaT}) and the linearized Einstein
equations define a closed 
set of integro-differential equations for the metric perturbations,
which may be solved by
a power-series ansatz \cite{Kraemmer}. 
Fig.~1(a) shows the comoving density contrast $\delta_{k}$ as a function
of conformal time $\eta$
of a perfect radiation fluid and a collisionless plasma respectively. 
At temperatures below the muon threshold ($T < 35$ MeV) the ratio 
$\alpha \equiv \rho_\nu/\rho_{\rm tot} \approx 0.49$,
whereas after $e^+ e^-$-annihilation 
at about $0.2$ MeV $\alpha \approx 0.41$. 
Fig.~1(b) shows the density contrasts for 
$\alpha = 1/2$. Between temperatures of $1$ MeV (decoupling of neutrinos)
and $10$ eV (the cold dark matter contributes only $1/10$ of the total 
energy density) this is a good model for the dominant matter perturbations
at that time. Neutrinos well above $1$ MeV are included in the 
radiation fluid usually. 
How can we describe neutrinos near the decoupling temperature and what 
effects should we expect for cosmological perturbations? 
\begin{figure}
\begin{center}
\vspace{-1.7 in}
\epsfig{figure=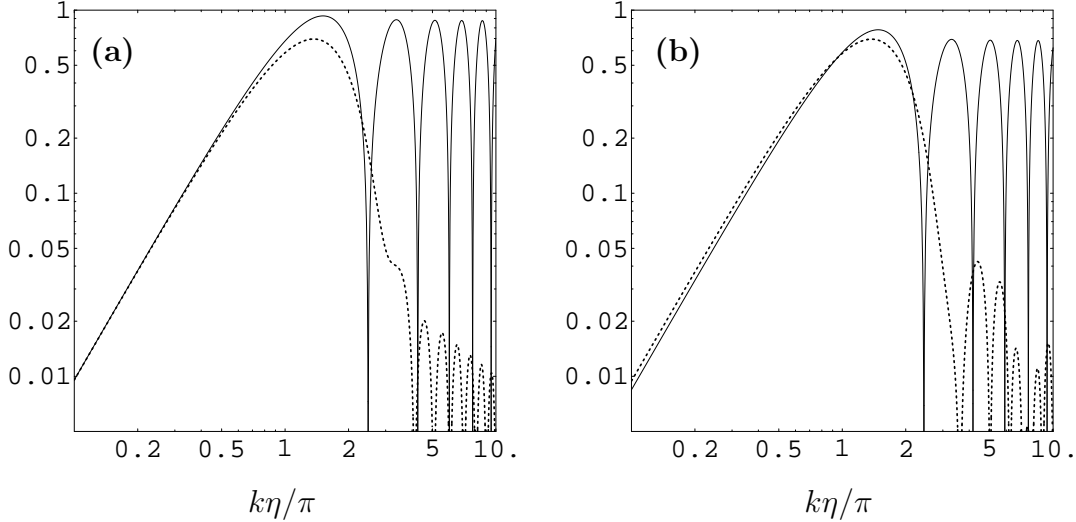,width=0.9\linewidth}
\vspace{-1.5 in}
\end{center}
\begin{center}
\vspace{-2.5 in}
\unitlength=1cm
\begin{picture}(12,6)
\put(0,6){\bf (a)}
\put(7.5,6){\bf (b)}
\put(2.1,0){$k\eta/\pi$}
\put(9.5,0){$k\eta/\pi$}
\end{picture}
\vspace{- 9pt}
\end{center}
\caption{{\bf (a)} The comoving density contrast 
grows as a function of the conformal time $\eta$ 
on superhorizon scales 
($k\eta \equiv k^{\rm phys}/H < \pi$) 
for the perfect fluid (full line)
and the collisionless plasma (dotted line). 
On subhorizon scales ($k^{\rm phys}/H > \pi$) the high sound speed, 
$c_s = 1/\sqrt{3}$, of the perfect fluid gives rise to acoustic oscillations of 
constant amplitude. Collisionless matter perturbations are 
damped due to directional dispersion, the phase speed is one, and the
oscillations are due to the gravitational interactions of the 
collisionless plasma with itself. 
{\bf (b)} A mixture of $50\%$ perfect radiation fluid (full line) and $50\%$
collisionless plasma (dotted line) evolved
together. On subhorizon scales the collisionless plasma feels the gravitational
potential of the radiation fluid, its damping is modulated. The normalization 
is arbitrary.}
\end{figure}

In a recent work H. Nachbagauer, A. Rebhan, and myself \cite{NRS}
studied the behavior of cosmological perturbations for thermal $O(N)$-symmetric
{\it scalar fields} with {\it weak self-interactions} $\lambda \phi^4$. 
We discovered an unexpected breakdown of
thermal perturbation theory on vastly subhorizon scales ($k^{\rm phys} \gg H$). 
The full set of 2-loop diagrams, after resummation of 
thermal masses $m=\lambda^{1/2}T + {\cal O}(\lambda)$
and nonlocal gravitational vertices, has been included 
in the gravitational polarization tensor of Eq.~(\ref{deltaT}). 
The resummation of an infinite subset of all
diagrams is standard in hot QCD \cite{Braaten}
and avoids infra-red divergences beyond the leading order $\lambda$. 
Due to the presence of a thermal mass the next-to-leading order 
is $\lambda^{3/2}$.
On superhorizon scales the behavior of the growing mode
is the same for any value of $\lambda$, but the decaying mode
may show superhorizon oscillations depending on the values of $\alpha$ and
$\lambda$ \cite{Kraemmer,NRS}. The subhorizon evolution of the 
density perturbation is shown in Fig.~2 for $\lambda = 1$.
\begin{figure}
\begin{center}
\vspace{-0.9 in}
\epsfig{figure=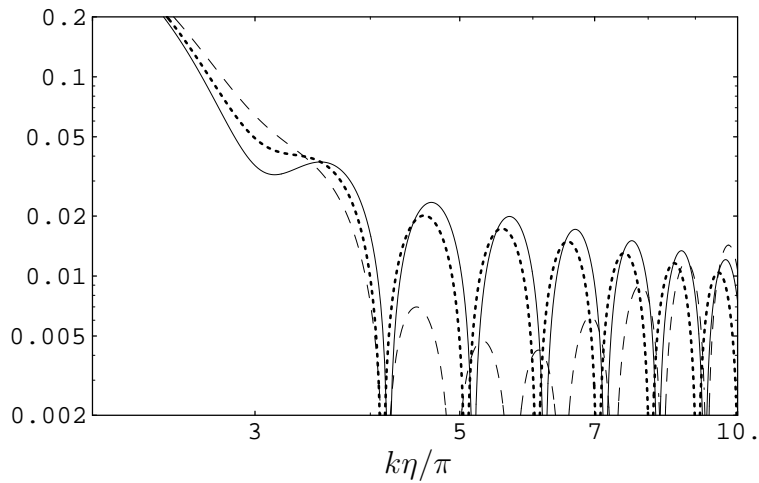,width=0.6\linewidth}
\vspace{-1 in}
\end{center}
\begin{center}
\vspace{-2.2 in}
\unitlength=1cm
\begin{picture}(12,6)
\put(6,0){$k\eta/\pi$}
\end{picture}
\vspace{- 9pt}
\end{center}
\caption{The subhorizon evolution of comoving density perturbations
of a collisionless plasma (dotted line as in Fig.~1(a)) 
is compared with the density perturbations of a self-interacting scalar 
plasma with $\lambda = 1$ (dashed line). At $k\eta/\pi \approx 4$ the 
thermal perturbation theory, evaluated through order $\lambda^{3/2}$,
starts to break down, which leads to unphysical growth of the perturbation
asymptotically. The subhorizon behavior of the solutions has been 
improved by use of Pad\'e approximants (full line). 
Vastly subhorizon the phase speed is smaller than one and the
damping is weaker than $1/(k\eta)$.}
\end{figure}
Thermal perturbation theory does not only break down for large values 
of the coupling $\lambda$, but also vastly subhorizon at 
perturbatively small values of $\lambda$.
Increasing orders in $\lambda^{1/2}$ are increasingly infra-red 
singular in the high
temperature limit as the external momentum $k$ approaches the light-cone.
For subhorizon scales $\eta \gg 1/k$
the contribution from $\lambda^1$ overtakes the contribution
from $\lambda^0$, $\lambda^{3/2}$ overtakes $\lambda^1$, and so on. 
A way to improve the subhorizon behavior is to rewrite the perturbative series
in $\lambda^{1/2}$ into a (2,1) Pad\'e approximant
\begin{equation}
a + b \lambda + c \lambda^{3/2} + \dots
\simeq a  + b \lambda\left/\left[1 - (c/b) \lambda^{1/2}\right]\right. + \dots
\end{equation}
In the large $N$ limit this gives an excellent approximation to the 
exact value of the thermal mass \cite{DJ} even for large values of $\lambda$. 
A calculation of the most important 1-loop diagram of the polarization tensor
to all orders 
in $\lambda^{1/2}$ has shown that the Pad\'e approximant, in contrast to the
perturbative result, 
describes the phases correct and provides a good approximation of the damping 
for large values of $\lambda$ and $\eta$. 
The full 2-loop Pad\'e improved solution for the density contrast with 
$\lambda = 1$ is shown in Fig.~2. The main conclusions for scalar
perturbations (for the vector and tensor sector see \cite{NRS}) are: \\
At superhorizon modes the anisotropic pressure decreases as the
coupling $\lambda$ is increased. \\
At subhorizon scales the phase speed of the damped oscillations is smaller 
than one, and \\
the damping is weaker than $1/(k\eta)$. \\
For one scalar field we have at subhorizon scales the paradoxical
situation that, although the interaction rate $\Gamma$ may be 
much bigger than the
Hubble rate, $\Gamma/H \sim \lambda^2 m_{\rm Planck}/T$, when 
$T\ll m_{\rm Planck}$ and $\lambda$ is perturbatively small, the 
plasma perturbations behave almost collisionless. This happens because 
the class of diagrams dropping out in the large $N$ limit 
(order $\lambda^2 \ln \lambda$ and higher) is 
not taken into account by the Pad\'e approximant solution. 
In the large $N$ limit, where the Pad\'e approximant is close to the exact
solution, the Hubble scale picks up a factor $\sqrt{N}$ from
the relativistic degrees of freedom, thus the interactions are slower
than the expansion. 
On superhorizon scales the interaction rate is irrelevant
for the behaviour of the perturbations. The amount of superhorizon anisotropic 
pressure is determined by the initial conditions and gravity only.

What does this tell us about neutrinos? At low temperatures $T \ll 100$ GeV
the weak interactions of neutrinos are well described by effective 
4-fermion interactions. The topology of the corresponding loop-diagrams 
is the same as before. Although the details are quite different 
the qualitative features should be similar. Let me naively relate
$\lambda$ to $G_F T^2$. At neutrino 
decoupling this corresponds to $\lambda \approx
10^{-11}$, thus perturbation theory should work perfect. Our results
for the scalar plasma suggest that neutrino perturbations behave almost
collisionless even above decoupling! On subhorizon scales this is not reliable
for two reasons: Again our result can only be trusted for a large number
of neutrino flavors, and the neutrinos couple to matter which is strongly 
interacting and is described by a perfect fluid. 
Our results show that, depending on initial conditions, 
effects of anisotropic pressure on superhorizon scales can be important
although the dominant plasma is collisional. 

I thank H. Nachbagauer and A. Rebhan for their collaboration on the issue
of this work. The work was supported in part by the FWF and the SNF.

\end{document}